\documentstyle[epsfig,prb,aps,twocolumn]{revtex}

\begin{document}
\title{Electrical detection of spin accumulation in a p-type GaAs quantum well.}
\author{R. Mattana, J.-M. George, H. Jaffr\`es, F. Nguyen Van Dau, A. Fert}
\address{Unit\'e Mixte de Physique CNRS/THALES, Domaine de Corbeville, 91404 Orsay, France, \\and Universit\'e Paris-Sud, 91405, Orsay, France}
\author{B. L\'epine, A. Guivarc'h, G. J\'ez\'equel}
\address{Equipe de Physique des Surfaces et Interfaces,
Unit\'e Mixte de Recherche CNRS-Universit\'e 6627 "PALMS",
Universit\'e Rennes I, 35042 Rennes Cedex, France}

\maketitle

\begin{abstract}
We report on experiments in which a spin-polarized current is
injected from a $GaMnAs$ ferromagnetic electrode into a $GaAs$
quantum well through an AlAs barrier. The resulting spin
polarization in the GaAs well is detected by measuring how the
current, tunneling to a second $GaMnAs$ ferromagnetic electrode,
depends on the orientation of its magnetization. Our results can
be accounted for the non-relaxed spin splitting of the chemical
potential, that is spin accumulation, in the $GaAs$ well. We
discuss the conditions on the hole spin relaxation time in GaAs
that are required to obtain the large effects we observe.
\end{abstract}


Introducing the spin as an additional degree of freedom in
semiconductor devices is an important challenge for the future of
spintronics\cite{prinz,wolf}. The semiconductors combine the
advantage of a long spin lifetime with the flexibility of their
carrier concentration and their high mobility. The long spin
coherence time in semiconductors have been evidenced by
time-resolved optical experiments and, for example, a spin
lifetime reaching a fraction of $\mu s$ has been evidenced in
n-doped $GaAs$ at low temperature\cite{kikkawa,dzhioev}. However,
the prerequisite of spin injection from a ferromagnetic conductor
in most concepts of devices raises difficult problems. It has
turned out that injecting spins from a ferromagnetic metal
encounters difficulties related to the conductivity mismatch
between metal and semiconductor\cite{schmidt,fertjaffres} and also
to their possible chemical incompatibility. This has driven the
development of magnetic semiconductors\cite{dietl,ohno,tanaka},
like $Ga_{1-x}Mn_{x}As$ which is ferromagnetic up to 110
K\cite{ohno}, more adapted for integration into semiconductor
heterostructures. Successful experiments on spin injection have
been achieved by injecting an electrical current from magnetic
semiconductors or metals and detecting the circular polarization
of emitted light\cite{ohno2,fiederling,jonker,zhu,young}. However,
if the principle is now well established, there is some questions
on how the efficiency of spin injection can be extracted from the
light polarization\cite{jansen}. In this letter we present
experiments of spin injection from a $GaMnAs$ electrode into a
$GaAs$ quantum well with detection of the polarization in $GaAs$
by measuring how the current, tunneling from $GaAs$ into a second
$GaMnAs$ electrode, depends on the orientation of its magnetic
moment. The structure is a double tunnel junction
$GaMnAs/AlAs/GaAs/AlAs/GaMnAs$. The first junction plays the role
of ballistic spin injector whereas the second one is used to
detect the spin accumulated in the semiconductor before being
transmitted. Our observation of large tunnel magnetoresistance
(TMR) effects demonstrates the efficient spin transmission across
$GaAs$. This is in contrast with the absence of spin transmission
in double junctions when the base is a nonmagnetic metal, and can
be explained by the non-relaxed spin polarization predicted for a
semiconductor base\cite{fertjaffres}.

Our double tunnel junctions, grown by molecular beam epitaxy on
semi-insulating $GaAs$ (001), are composed of two ferromagnetic
electrodes ($Ga_{1-x}Mn_{x}As$) separated by a
$AlAs(1.5nm)/GaAs(5nm)/AlAs(1.5nm)$ trilayer. Thin layers of GaAs
(1nm) are also intercalated between the $GaMnAs$ and $AlAs$ layers
to prevent interdiffusion between the two materials. To probe the
spin-polarization of electron tunneling from $GaMnAs$ through
$AlAs$, test experiments have been also performed on single tunnel
junctions where the central trilayer of the double junction is
replaced by a single 1.7 nm thick $AlAs$ barrier\cite{thickness}.
Structures have been deposited at $230^{o}C$ on a GaAs buffer
layer grown at $580^{o}C$. Junctions with diameter from $10\mu m$
to $300\mu m$ were patterned by optical
lithography\cite{montaigne}. Ohmic contacts on both $GaMnAs$
electrodes were made by deposition of Ti ($50nm$) and Au
($150nm$).

Different thicknesses and $Mn$ concentrations have been chosen for
the two electrodes in order to obtain different coercive fields
and then an antiparallel magnetic configuration. The bottom and
top $Ga_{1-x}Mn_{x}As$ films have respective thicknesses of
$300nm$ and $30nm$. The $Mn$ concentration is 4.3\% and 5.3\%
(bottom and top electrode) for the double barrier structure, 4.7\%
and 5.4\% (bottom and top) for the single barrier. M(H) hysteresis
loops of the heterostructures before patterning show two steps
associated to the reversal of the two $GaMnAs$ layers at different
coercive fields. The remanent magnetization is only 30\% of the
saturated magnetization which is reached at about 1 Tesla. The
magnetization of the sample collapses near 50 K (Curie
temperature) and the absence of remanent magnetization above this
temperature indicates there is no formation of $MnAs$ clusters.
Higher Curie temperatures than 50K have been obtained for
$Ga_{1-x}Mn_{x}As$ with $x\approx 5\%$ after thermal treatments.
Nevertheless, we have not annealed our junctions to avoid possible
diffusions into the $AlAs$ barriers. We have however checked the
TMR of our single junctions (a probe of the spin polarization) is
nearly as high ($38 \%$) as for the junctions with the same $AlAs$
thickness in Ref.[9]

\begin{figure}
\centering
\epsfig{file=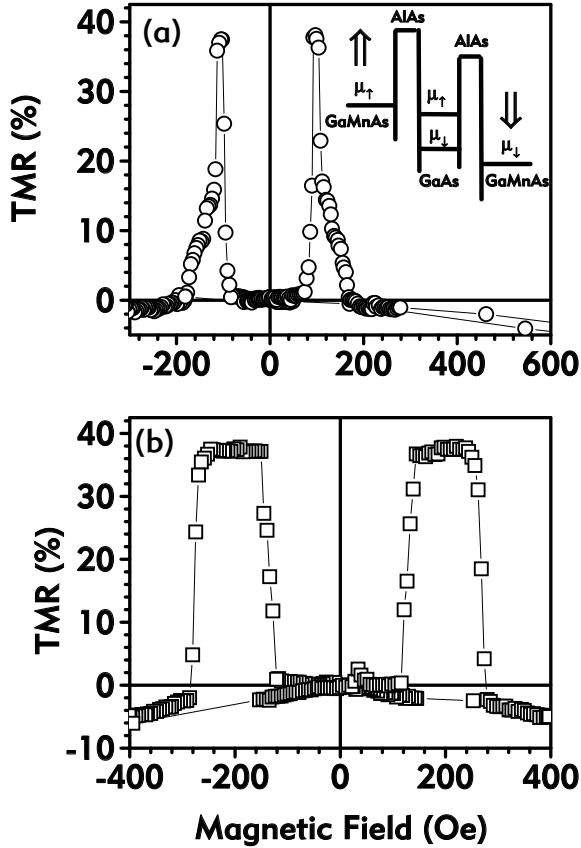, width=8cm} \caption{(a) TMR curve for a
$20\mu m$ diameter double barrier junction. Inset of
Fig.\ref{fig:fig1}a : Schematic picture of the spin splitting of
the electro-chemical potentials $\mu_{\uparrow}$ and
$\mu_{\downarrow}$ in the nonmagnetic central layer of a F/I/N/I/F
structure in the antiparallel state(from Ref.\cite{fertjaffres}).
For convenience, the picture for holes has been translated into a
picture for electrons. (b) TMR curve for a single barrier MTJ. In
both (a) and (b), the measurements have been recorded at 4K and at
a bias of 1mV. The magnetic field was applied along [100] axis.
}\label{fig:fig1}
\end{figure}

In Fig.\ref{fig:fig1}, we show the TMR of the double barrier
(Fig.\ref{fig:fig1}a) and single barrier (Fig.\ref{fig:fig1}b)
junctions at 4 K. In both cases, the magnetic field is set along
the [100] magnetic easy axis and the TMR is derived from
four-contact measurements at constant bias voltage (1mV). The TMR
($\Delta R/R_{0}$ where $R_{0}$ is the zero field resistance) is
associated with the switching between the parallel (P) and
antiparallel (AP) configurations of the remanent magnetizations
($30\%$ of saturation). Similar TMR results on single barrier
junctions have been found by Tanaka and Higo\cite{tanaka}. For a
thickness of 1.7 nm for $AlAs$, these authors find a TMR ratio
$\cong 45\%$ that is approximately the same as what we measure on
single barrier structure.

\begin{figure}
\centering
\epsfig{file=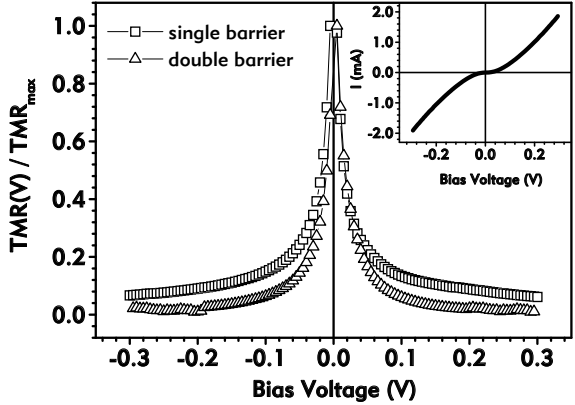, width=8cm} \caption{Bias dependence of TMR
for double barrier (triangles) and a single (squares)
  junctions. Inset: I(v) curve at zero field measured
  at 4K on the double barrier junction.}\label{fig:fig2}
\end{figure}

Figure \ref{fig:fig2} shows that the TMR amplitude decreases
rapidly with the bias voltage. This bias dependence of the TMR is
derived from the difference between I(V) curves (inset of Fig.2)
recorded at zero field and in the antiparallel magnetic
configuration. The bias dependence of Fig.2 is confirmed by R(H)
curves recorded at different bias.

A striking result is that the double junctions exhibit TMR effects
(at a level similar to that of the single junction), in contrast
to which could be expected for F/I/N and N/I/F junction in series.
In such F/I/N/I/F double barrier where N is nonmagnetic, TMR is
expected in the following cases:

a) Hot electrons having not relaxed their energy and transmitted
above the second barrier\cite{monsma,Dijken}. This can produce a
significant TMR for applied voltage exceeding the barrier height
but not in the very small voltage limit of our experiments.

b) Predominant direct tunneling between the ferromagnetic
electrodes through the entire $AlAs/GaAs/AlAs$ barrier would give
TMR effects but also a much higher tunnel resistance. According
to the results of Tanaka and Higo\cite{tanaka}, increasing the
thickness of AlAs from 1.7 nm (thickness in our single barrier
junction) to 3.4 nm (double barrier) would increase the
resistance by almost four orders of magnitude and, in addition,
decrease the TMR. We can rule out direct tunneling in our double
junctions because their resistance and TMR are both close to
those of the single junction (around $10^{-2}\Omega.cm^{2}$ for
the resistance and $38\%$ for the TMR).

c) Coherent resonant tunneling on quantum well states in $GaAs$
would give TMR effects but would be characterized by a specific
bias dependence of the conductance (a negative differential
conductance for example)\cite{hayden} and TMR. We have plotted on
Fig.\ref{fig:fig2} the bias dependence of the TMR and conductance
in inset. We do not observe any fine structure that could be the
signature of a coherent resonance on discrete levels. Actually, a
coherent resonant tunneling would require a coherence time of the
wave functions in the well, $\tau_{c}$, longer than the mean time
spent by the hole in the well, $\tau_{n}$. This condition will be
discussed later.

d) Sequential tunneling without spin relaxation in the $GaAs$ well
(or more generally in a semiconducting spacer) can give TMR
effects, as calculated in Ref.\cite{fertjaffres}. This process is
expected to produce a large TMR if the spin relaxation time
$\tau_{sf}$ is larger than the mean time $\tau_{n}$ spent
ballistically in the spacer between the two successive tunnelings
(in principle, the TMR of the double junctions should be half that
of the simple ones\cite{fertjaffres}, but, as our double junctions
have slightly thinner $AlAs$ barriers, the increase of the
spin-polarization of tunneling at dcreasing thickness\cite{tanaka}
should balance more or less the reduction by a factor of 2). For
the situation of diffusive transport in the spacer, this
corresponds to the second condition of Eq.35 in
Ref.\cite{fertjaffres}, that is $r_{T} \ll \rho_{N}
l_{sf}^{2}/t_{N}$ or equivalently $t_{N} \ll \rho_{N}
l_{sf}^{2}/r_{T}$ ($t_{N}$=spacer thickness, $\rho_{N}$ and
$l_{sf}^{N}$ = resistivity and spin diffusion length in the
spacer, $r_{T}$=tunnel resistance). To our knowledge, such
sequential tunneling with spin conservation has never been
observed up to now.

We ascribe the TMR of our double junction to the mechanism d).
This is also supported by the discussion below on the three
characteristic times $\tau_{n}$, $\tau_{c}$ and $\tau_{sf}$.

The time $\tau_{n}$ is related to the broadening of the quantized
energy level $\epsilon_{n}$ and can be expressed as a function of
$\epsilon_{n}$ and $\bf{T}$ -the transmission coefficient of the
detection tunnel barrier- by $\tau_{n}=\pi
\hbar/(\epsilon_{n}\bf{T})$\cite{gueret}. This expression can be
directly derived from the picture of holes reflecting $2/\bf{T}$
times against the barriers with a kinetic energy $\epsilon_{n}$
before being ejected out of the well. A typical energy of some
tens of $meV$ for a few nm thick well\cite{hayden,wessel} results
in a value of $\tau_{n}$ of the order of 100 $ps$ for a
transmission coefficient $\bf{T}$ of the order of $10^{-3}$ (this
is the value derived from the variation of the tunnel resistance
as a function of the barrier thickness in the experimental
results of Tanaka and Higo\cite{tanaka}).

With a value of the order of 100 ps, $\tau_{n}$ is much longer
than the coherence time $\tau_{c}$ ($\cong$ inelastic relaxation
time, which is generally less than 1 ps). This shows that the
condition for resonant tunneling is not fulfilled and this
implies that we are in a regime of sequential tunneling. In this
regime, we expect that the double junction exhibits a TMR at the
level of the TMR of the single junction if the spin relaxation
time $\tau_{sf}$ is much longer than $\tau_{n}$, that is 100 ps
and thus approaches the ns range. This seems to be in agreement
with the results of optical measurements on hole spin lifetime in
$GaAs$ quantum well\cite{roussignol}. This enhancement of the hole
spin lifetime at low temperature compared to the bulk value,
measured recently at $\simeq 100 fs$\cite{hilton}, can be
understood as the effect of i) the lift of the valence-band
degeneracy between the $J_{z}=\pm 3/2$ and $J_{z}=\pm 1/2$ states
at the $\Gamma$ point produced by the confinement or equivalently
ii) the strong reduction of the solid angle of hole wave-vectors
around the quantized direction (that is a small $k$ parallel
component) as $k_{B}T$ remains small compared with
$\epsilon_{n}$\cite{roussignol}.

The condition $\tau_{sf}\ll \tau_{n}$ of the discussion above can
be related to the condition expressed in the model of
Ref.\cite{fertjaffres}, that is in a picture with a splitting of
the spin up and spin down electro-chemical potentials (Fermi
energies) in the AP configuration (as illustrated in the inset of
Fig.1a). This splitting simply reflects that, in the AP
configuration, one injects a majority of spin up holes whereas a
majority of spin down holes tunnel towards the outer electrode,
thus generating an imbalance between the two populations of spin.
If the resulting spin splitting $\Delta \mu$ does not relax from
its maximum value (of the order of the total voltage drop $V$
between ferromagnetic electrodes), this gives rise to the same
TMR as in the single junctions. The condition of negligible
relaxation is having a number of spin flips per unit of time and
unit area in the well, that is $(\Delta
\mu/k_{B}T)n^{2D}/\tau_{sf}$ at small bias, much smaller than the
injected spin current of the order of $j/e$ ($n^{2D}$ is the
density of the 2-Dimensional gaz -2-DEG- in the
well)\cite{fertjaffres}. Expressing $\tau_{sf}$ as a function of
the spin diffusion length $l_{sf}$ in the well and hole mobility
$\nu$ from the relation $l_{sf}=\sqrt{k_{B}T \nu
\tau_{sf}/e}$\cite{remark1}, the condition for maintaining
$\Delta \mu$ at a level of the order of $eV\simeq er_{T}J$
($r_{T}$ is the tunnel resistance) can be written as :
\begin{equation}
\label{condition} r_{T}\ll \frac{l_{sf}^{2}}{n^{2D}e\nu}
\end{equation}
which is the 2DEG version of the condition $r_{T} \ll \rho_{N}
[l_{sf}^{N}]^{2}/t_{N}$ in Ref.\cite{fertjaffres}. The
equivalence of Eq.\ref{condition} with the condition $\tau_{n}\ll
\tau_{sf}$ turns out straightfully if the tunnel resistance is
related to the transmission coefficient $\textbf{T}$ by a
Landauer-like formula, $r_{T}=\alpha h/e^{2}$ $n_{2D}^{-1}$
$\textbf{T}^{-1}$\cite{Landauer,datta} and then to $\tau_{n}$ by
$\tau_{n}=\pi \hbar/(\epsilon_{n}\bf{T})$ (the coefficient
$\alpha=k_{B}T/\epsilon_{n}$ is a reduction factor expressing
that the energy $\epsilon_{n}$ of the confined state is larger
than $k_{B}T$).

To probe our interpretation, we have also measured the TMR of a
double junction with the same value of $t_{N}$ ($GaAs$ spacer
thickness) but with a higher value of $r_{T}$ (by about a factor
of 10). The TMR curves are very similar to those of
Fig.\ref{fig:fig1} but the amplitude is only $3\%$ instead of
$38\%$. This is consistent with the reduction of the
spin-polarization when $r_{T}$ becomes too large to satisfy
Eq.\ref{condition}. In other words, this means that, with the
smaller transmission coefficient \textbf{T} associated to a
higher $r_{T}$, the time spent by the hole in the $GaAs$ well,
$\tau_{n}=\pi \hbar/(\epsilon_{n}\bf{T})$, is no longer much
smaller than the spin relaxation time $\tau_{sf}$ and the spin
relaxation in $GaAs$ reduces the TMR.

The spin splitting of the electro-chemical potential in a
nonmagnetic spacer between two tunnel junctions has already been
detected by Jedema \textit{et al.}\cite{jedema}, this time for a
metallic spacer ($Cu$) in $Co/Al_{2}O_{3}/Cu/Al_{2}0_{3}/Co$
double junction structures. However, in this case, the splitting
is much smaller than the potential drop between the magnetic
electrodes, typically 10 $\mu eV$ compared to the potential drop
of the order of 100 $meV$. This can be expected from
Eq.\ref{condition} or the equivalent condition for a 3D-spacer,
$r_{T}<\rho^{3D}l_{sf}^{2}/t_{N}$, where $\rho^{3D}$ is the
resistivity of the spacer. Actually, with the typical
low-resistivity of metals (low compared to semiconductors) and
spin diffusion length $l_{sf}$ in the micron range, the above
condition for obtaining a spin splitting of the order of the
potential drop across the double junction would require to have
$r_{T}$ not higher than $0.1 \Omega \mu m^{2}$. With resistances
of alumina barriers of the order of $1 k\Omega \mu m^{2}$, the
spin splitting turns out to be a very small fraction of the total
potential drop across the double junction. More generally, this
also explains that a significant TMR could never been observed in
double junctions in which the central layer is a nonmagnetic metal
with such high tunnel resistance. The situation with a
semiconductor central layer of high resistance is much more
favorable.

In conclusion, we have presented experiments in which, after
injection of a spin-polarized current into a $GaAs$ quantum well
from a $GaMnAs$ electrode, the spin polarization in $GaAs$ is
detected by measuring the spin polarization of the current
tunneling into a second $GaMnAs$ electrode. We have shown that our
results can be explained by sequential tunneling with low enough
spin relaxation in the $GaAs$ layer. The TMR of our double
junction ($38\%$) has the same order of magnitude as the TMR of
the single ones single ones, what can be expected if the condition
of negligible relaxation in $GaAs$, Eq.\ref{condition}, is
satisfied. We have also shown that this condition can be hardly
satisfied with a metallic spacer instead of $GaAs$, so that the
effects we observe are specific of spin injection into
semiconductors. To our knowledge, these experimental results
represent the first clear evidence of an electrical spin detection
of spin injection into a semiconductor. Further experiments on
similar structures with various thicknesses of the central well
(or layer) or various dopings and carrier densities should lead to
a more general understanding of the conditions for spin injection
and electrical spin detection in semiconductors. We point out that
injection into n-type semiconductors having a larger spin lifetime
should allow spin propagation on longer distances. Injecting spins
from a third contact for an additional control of the spin
polarization should lead to new types of spintronic devices.

This research was supported by a French program Action
concert\'ee Nanosciences-Nanotechnologies and by the EU through
the RTN 'Computational Magnetoelectronics' (HPRN-CT-2000-00143).


\begin{thebibliography}{10}

\bibitem{prinz}
G. Prinz, Science {\bf 282},  1660  (1998).

\bibitem{wolf}
S.~A. Wolf {\it et~al.}, Science {\bf 294},  1488  (2001).

\bibitem{kikkawa}
J.M. Kikkawa and D.D. Awschalom, Phys. Rev. Lett. {\bf 80}(19),
4313-4316 (1998)

\bibitem{dzhioev}
R.I. Dzhioev {\it et~al.}, JETP Lett. {\bf 74}(3), 182-185 (2001)

\bibitem{schmidt}
G. Schmidt {\it et~al.}, Phys. Rev. B {\bf 62},  4790  (2000).

\bibitem{fertjaffres}
A. Fert and H. Jaffres, Phys. Rev. B {\bf 64},  184420  (2001).

\bibitem{dietl}
T. Dietl, H. Ohno, F.Matsukura, J. Cibert and D. Ferrand, Science
{\bf 287}, 1019 (2000).

\bibitem{ohno}
H. Ohno, J. Magn. Magn. Mat. {\bf 200}, 110  (1999).

\bibitem{tanaka}
M. Tanaka and Y. Higo, Phys. Rev. Lett. {\bf 87},  026602 (2001);
M. Tanaka and Y Higo, Physica E {\bf 10}, 292 (2001); Y. Higo, H.
Shimizu and M. Tanaka, J. Appl. Phys. {\bf 89}, 6745 (2001).

\bibitem{ohno2}
Y. Ohno {\it et~al.}, Nature {\bf 402},  790  (1999).

\bibitem{fiederling}
R. Fiederling {\it et~al.}, Nature {\bf 402},  787  (1999).

\bibitem{jonker}
A. T. Hanbicki and B. T. Jonker and G. Itskos and G. Kioseoglou,
and A. Petrou Appl. Phys. Lett. {\bf 80}, 1240 (2002)

\bibitem{zhu}
H. J. Zhu, M. Ramsteiner, H. Kostial, M. Wassermeier, H.-P.
Schönherr and K. H. Ploog, Phys. Rev. Lett. {\bf 87}, 016601
(2001)

\bibitem{young}
D.K. Young, E. Johnston-Halperin, D.D. Awschalom, Y. Ohno and H.
Ohno, Appl. Phys. Lett. {\bf 80}(9), 1598-1600 (2002)

\bibitem{jansen}
see comments of R. Jansen, Appl. Phys. Lett. {\bf 81}(11), 2130
(2002); A.T. Hanbicki and B.T. Jonker, Appl. Phys. Lett. {\bf
81}(11), 2131 (2002)

\bibitem{thickness}
1.7 nm is in agreement with cross-section TEM images of the single
junctions. Although the double barriers were grown to obtain the
same $AlAs$ thickness, their resistance R is smaller (by a factor
about 2, that is by a factor of approximately 4 for each
individual barrier). The comparison with the R(t) curves of
Ref.[9] indicates that $t_{AlAS}\simeq $1.5nm. From the same
reference, the TMR ratio is about 70 \% for 1.5 nm and 45
\% for 1.7nm.


\bibitem{montaigne}
F. Montaigne {\it et~al.}, Appl. Phys. Lett. {\bf 73},  2829
(1998).

\bibitem{tanaka2}
T. Hayashi, M. Tanaka and A. Asamitsu, J. Appl. Phys. {\bf 87},
4673 (2000).

\bibitem{monsma}
D.~J. Monsma, J.~C. Lodder, T.~J.~A. Popma, and B. Dieny, Phys.
Rev. Lett. {\bf 74},  5260  (1995).

\bibitem{Dijken}
S. van Dijken, X. Jiang, and S.~S.~P. Parkin, Appl. Phys. Lett.
{\bf 80},  3364 (2002).

\bibitem{gueret}
P. Gu\'eret and C. Rossel, \textit{'Resonant Tunneling in
Semiconductors. Physics and Applications'}, Plenum Press, Nato ASI
Series {\bf 277} p.82, Eds. L.L. Chang, E.E. Mendez and C.
Tejedor (1991)

\bibitem{hayden}
R.K. Hayden {\it et~al.} Phys. Rev. Lett. {\bf 66}, 1749-1752
(1991)

\bibitem{wessel}
R. Wessel and M. Altarelli, Phys. Rev. B, {\bf 39}, 12802-12807
(1989)

\bibitem{ohno3}
Y. Ohno, I. Arata and F. Matsukura and H. Ohno, Physica E, to be
published.

\bibitem{roussignol}
Ph. Roussignol {\it et~al.}, Phys. Rev. B {\bf 46}, 7292-7295
(1992); S. Bar-Ad and I. Bar-Joseph, Phys. Rev. Lett. {\bf 68},
349-352 (1992)

\bibitem{hilton}
D.J. Hilton and C.L. Tang, Phys. Rev. Lett. {\bf 89}, 146601
(2002)

\bibitem{remark1}
from $l_{sf}\propto \sqrt{\lambda \lambda_{sf}}$ where
$\lambda=v_{th}\tau$ and $\lambda_{sf}=v_{th} \tau_{sf}$ are
respectively the mean free path and the spin mean free path in
the well and $v_{th}$ the thermal velocity.

\bibitem{Landauer}
R. Landauer, IBM J. Res. Dev., {\bf 1}, 223 (1957);

\bibitem{datta}
S. Datta, \textit{Electronic Transport in Mesoscopic Systems}
(Cambridge University Press, Cambridge, England, 1995), pp.
151-157

\bibitem{jedema}
F.~J. Jedema {\it et~al.}, Nature {\bf 416},  713  (2002).

\end{thebibliography}
\end{document}